\documentclass[3p,times]{elsarticle}
\usepackage{algorithm}
\usepackage{algorithmic, amsmath}
\usepackage{epsfig,graphicx}
\usepackage{enumerate}
\usepackage{calligra}
\usepackage{mathptmx}       
\usepackage{helvet}         
\usepackage{courier}        
\usepackage{type1cm}        
\usepackage{makeidx}         
\usepackage{graphicx}        
\usepackage{multicol}        
\usepackage[bottom]{footmisc}
\usepackage{adjustbox}
\usepackage{xcolor}

\begin{document}
\begin{frontmatter}
\cortext[cor1]{Corresponding Author}
\title{Efficient  Edge Rewiring Strategies for Enhancement in Network Capacity}
\author{Suchi Kumari, Abhishek Saroha, Anurag Singh$^*$}
\address{Department of Computer Science \& Engineering, National Institute of Technology, Delhi, India}
\ead{suchisingh@nitdelhi.ac.in,abhishek.saroha67@gmail.com,anuragsg@nitdelhi.ac.in}

\begin{abstract}
The structure of the network has great impact on its traffic dynamics. Most of the real world networks follow heterogeneous structure and exhibit scale-free feature. In scale-free network, a new node prefer to connect with hub nodes and the network capacity is curtailed by smaller degree nodes. Therefore, we propose rewiring a fraction of links in the network, to improve the network transport efficiency. In this paper, we discuss some efficient link rewiring strategies and perform simulations on scale-free networks, confirming the effectiveness of these strategies. The rewiring strategies actually reduce the centrality of the nodes having higher betweenness centrality. After the link rewiring process, the degree distribution of the network remains the same. This work will be beneficial for the enhancement of network performance.
\end{abstract}

\begin{keyword}
Rewiring strategies \sep centrality \sep traffic capacity
\end{keyword}

\end{frontmatter}

\section{Introduction}
Traffic dynamics of the time-varying networks has attracted the attention of a large number of researchers in recent years \cite{danila2007,xue2010,chen2018}. The ever-growing demand for the network resources has ineluctably led to traffic congestion in many real-life networks such as communication, transportation and so on. Most of the network structures are evolving and scale-free in nature where hubs and small degree nodes coexist. Congestion usually initiates from these hubs and then gradually propagates in the whole network. Here, we propose some efficient rewiring strategies to maximize the traffic capacity of the hub nodes and the network performance as well.  

The structure of the network has great impact on the traffic capacity, especially the communication network in the era of bursty traffic generation. Therefore, the research of the enhancement of the network capacity is vital to us and has been studied extensively. The network capacity can be improved either by implementing various routing strategies \cite{yan2006,danila2007} or by optimizing the underlying network structure \cite{xue2010,chen2018}. Implementation of optimal routing strategies is comparatively easier and incurs a low cost. However, finding of a global optimal route is an NP-complete problem \cite{bui}. Although network restructuring incurs more cost, the effect of varying network structure on the traffic capacity and the load at the network is essential for the study of traffic dynamics on complex networks. 

Network structure may be optimized by redistributing the traffic of heavily loaded \cite{kumari2017}, removing some congested nodes or links \cite{jiang2014}, adding some network resources \cite{huang2010} and so on. In real networks, the resources such as capacity, queue size are finite. Addition of new resources incur cost hence, rewiring strategies may be more suitable to improve network capacity. Rewiring strategies depend on numerous network parameters such as degree, centrality, queue length etc. Jiang \textit{et al.} \cite{jiang2014} improved the network capacity by removing the connection between two nodes with the maximum sum of the degree together with betweenness centrality (BC) and by adding a connection between two unconnected nodes with the minimum sum of the degree in addition to BC. Alweimine \textit{et al.} \cite{alweimine} prioritized the packets according to the destination node and found that the prioritization of nodes with hubs is always more efficient than the prioritization of nodes with a small degree or a random prioritization. Ma \textit{et al.} \cite{ma} worked on a two-layer network model and proposed a delivery capacity allocation strategy based on the degree distributions of both the layers.  Jiang \textit{et al.} \cite{jiang2013optimal}, assigned capacity dynamically to each link proportional to the queue length of the link. Larger delivery capacities can be assigned to the hub nodes and a tunable parameter is introduced to optimize network capacity. 

All the above researchers considered degree and centrality as the network parameters for rewiring and removal of links and nodes as well. Assortativity (or disassortativity) is another network parameter which measures the tendency of a node to connect with other node having similar (or dissimilar) attributes, such as the degree. It can also be used to enhance traffic capacity based on similarity or dissimilarity of an individual node. Fronczak \cite{fronczak} discussed the effect of assortativity on the properties of transport in complex networks and calculated critical packet generation rate $ \lambda_c $ in traffic based on routing strategy with local information. Xue \textit{et al.}\cite{xue2010} tuned the transport efficiency on scale-free networks through the disassortative or assortative topology. Chen \textit{et al.}\cite{chen2018} rewired the link against traffic congestion and proved that the network should have a core-periphery structure and disassortative in nature. 

Networks can be represented and analyzed by using local, global and intermediate (meso) level perspectives. Degree and assortativity (or disassortativity) are calculated on local level. Centrality measures such as BC, closeness centrality (CC) etc. are computed globally.  In the meso level, we focus on a group of nodes instead of focusing on an individual node or the whole in the network. With the help of intermediate level parameters, we may discover some important features those are not covered either at local level or global level (e.g., summary states of algorithmic identification).  Core-periphery structure \cite{rombach} is a kind of meso level representation in which some nodes constitute a densely connected core and some make up the sparsely connected periphery. Apart from all these parameters, rich club phenomenon is also a network parameter and its presence increases the load and congestion on the link between two hubs. Inspired by the importance of all these parameters, we have considered all the perspectives for the restructuring and performance measurement (in terms of traffic capacity) of the network. The key contributions of the proposed work are as follows. (i) Generate the optimal network structure for maximizing the network capacity using various rewiring strategies in scale-free networks. (ii) Provide a traffic flow model to estimate actual number of packets in the network. (iii) Analyze the networks by studying several local and global network parameters, e.g., maximum betweenness centrality, clustering coefficient, average path length (APL), assortativity, rich-club phenomenon, core-periphery structure. (iv) Establish a relationship between the network parameters and the traffic performance of the network by fixing its degree distribution.

Section 2 provides some basic definitions of various parameters. In Section 3, some rewiring strategies are described. Section 4 discusses about the traffic flow models. Section 5 presents the simulation results, and in the last section, conclusions and future research plan are discussed.
\section{Definitions}
\hspace{-0.65 cm} \textbf{Centrality measures}\\
\smallskip
\textit{Betweenness centrality (BC)} is used to measure the extent to which a node lies on shortest paths between other node pairs. Betweenness centrality, $ g(v) $of a node $ v $ is equal to the number of shortest paths from all node pairs pass through that
node $ v $ and is given by $ \sum_{s \neq v \neq d} \frac{\sigma_{s \rightarrow d}(v)}{\sigma_{s \rightarrow d}} $, where $ \sigma_{s \rightarrow d} $ is the total number of shortest paths from node $ s $ to node $ d $ and $ \sigma_{s \rightarrow d}(v) $ is the number of those paths that pass through node $ v $.  \textit{Closeness centrality (CC)} of a node $ u $ is the reciprocal of sum of the shortest path distances from all other nodes $ V\setminus u $ to $ u $ within the network and it is normalized by the sum of all possible short distance i.e., $
\mathcal{C}(u) = \frac{N-1}{\sum_{v\neq u \in V} d(v, u)} $
where, $ d(v,u) $ is the shortest path distance between $ v $ and $ u $ and $ N $ is the total number of nodes  and $ V $ is the set of nodes in the network. \textit{Eigenvector centrality (EC)} is used to find impact of one node to others. EC of node $ i $, $ \hat{x}(i) = \frac{1}{\kappa} \sum_{j \in V} A_{i,j} \hat{x}(j) $. $ \kappa $ is a constant value, $ A $ is the adjacency matrix and the value of $ A_{i,j} = 1 $ if there is a link between the node $ i $ and the node $ j $, else it is $ 0 $.

\smallskip
\hspace{-0.5 cm}\textbf{Degree-degree correlation}\\
Degree-degree correlation (DDC) exhibits the relationship between the degrees of each pair of nodes that connect with each other. Let $ p_k $ is the probability of a randomly chosen node having degree $ k $. After reaching a node $ i $, the degree distribution of the reached node is called as excess degree distribution; $ q_k = \frac{(k+1) p_{k+1}}{\langle k \rangle} $. The value of network's correlation coefficient, $ r_{deg} $ can be written in the form of correlation function as $ r_{deg} = \langle jk \rangle- \langle j \rangle \langle k \rangle = \sum_{jk} jk(p_{e_{jk}}-q_jq_k) $. Here, $ p_{e_{jk}} $  is the joint degree distribution of a randomly chosen link, $ e_{jk} $ of the two nodes at either end of the link and $ \langle k \rangle $ is average degree of the network. The normalized value of correlation coefficient, $ r_{deg} $ is defined as \cite{newmanassortative},
\begin{equation}
r_{deg} = \frac{1}{\sigma (q)^2} \sum_{j,k} jk\{ p_{e_{j,k}}- q_jq_k \}. \label{e1}
\end{equation}
Here, $ \sigma (q)^2 = \sum_k k^2 q_k - {[\sum_k k q_k]}^2 \mbox{ and } -1 \leq r_{deg} \leq 1 $. The value of $ r_{deg} $ is positive (or negative) for assortative (or disassortative) network while zero for uncorrelated networks. DDC between two nodes $i $ and $ j $ can be defined as  $ r_{deg}(i,j) = \frac{cov(i,j) } { \sqrt{ cov(i,i) * cov(j,j) } } $, where, $ cov $ is the covariance matrix.

\begin{figure}[h]
\centering
\begin{minipage}[b]{0.45\linewidth}
\centering
$\begin{array}{cc}
\includegraphics[width=0.45\linewidth, height=0.15\textheight]{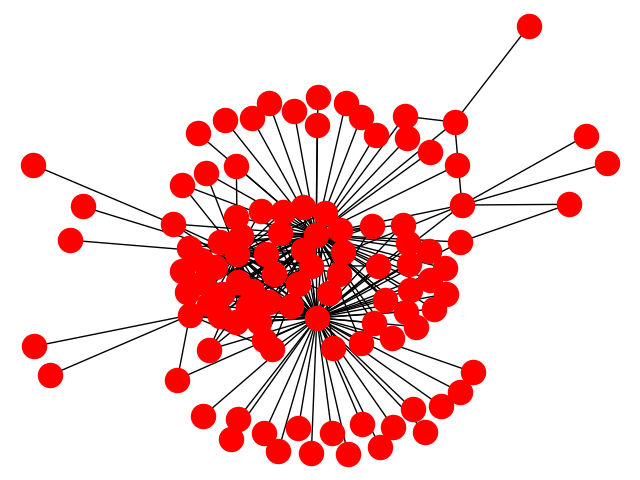} &
\includegraphics[width=0.45\linewidth, height=0.15\textheight]{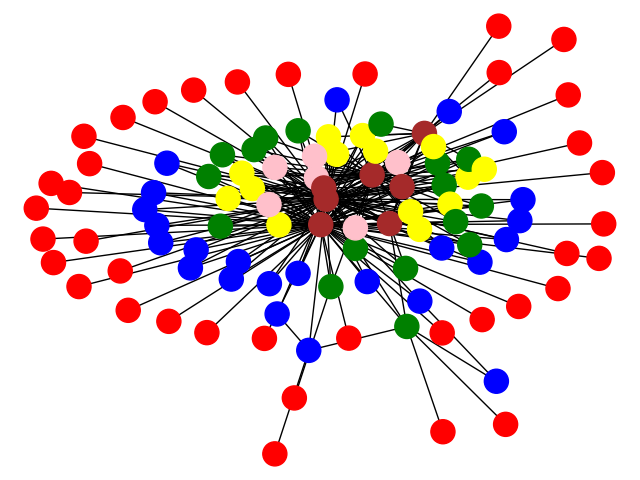}\\
\mbox{(a) Original Network} & \mbox{(b) } k\mbox{-core architecture}
\end{array}$
\caption{Visualization of the core-periphery structure of a European airline network ($ N =106 $). The network has several layers of cores and a small dense innermost core. Best viewed in the various color schemes. Red, blue, green, yellow, pink and brown color represent subnetwork with $ 1 $-core, $ 2 $-core, $ 3 $-core and $4$-core, $ 5 $-core and $ 6 $-core respectively.}
\label{f1}
\end{minipage}
\hspace{0.15cm}
\begin{minipage}[b]{0.45\linewidth}
$\begin{array}{cc}
\includegraphics[width=0.45\linewidth, height=0.15\textheight]{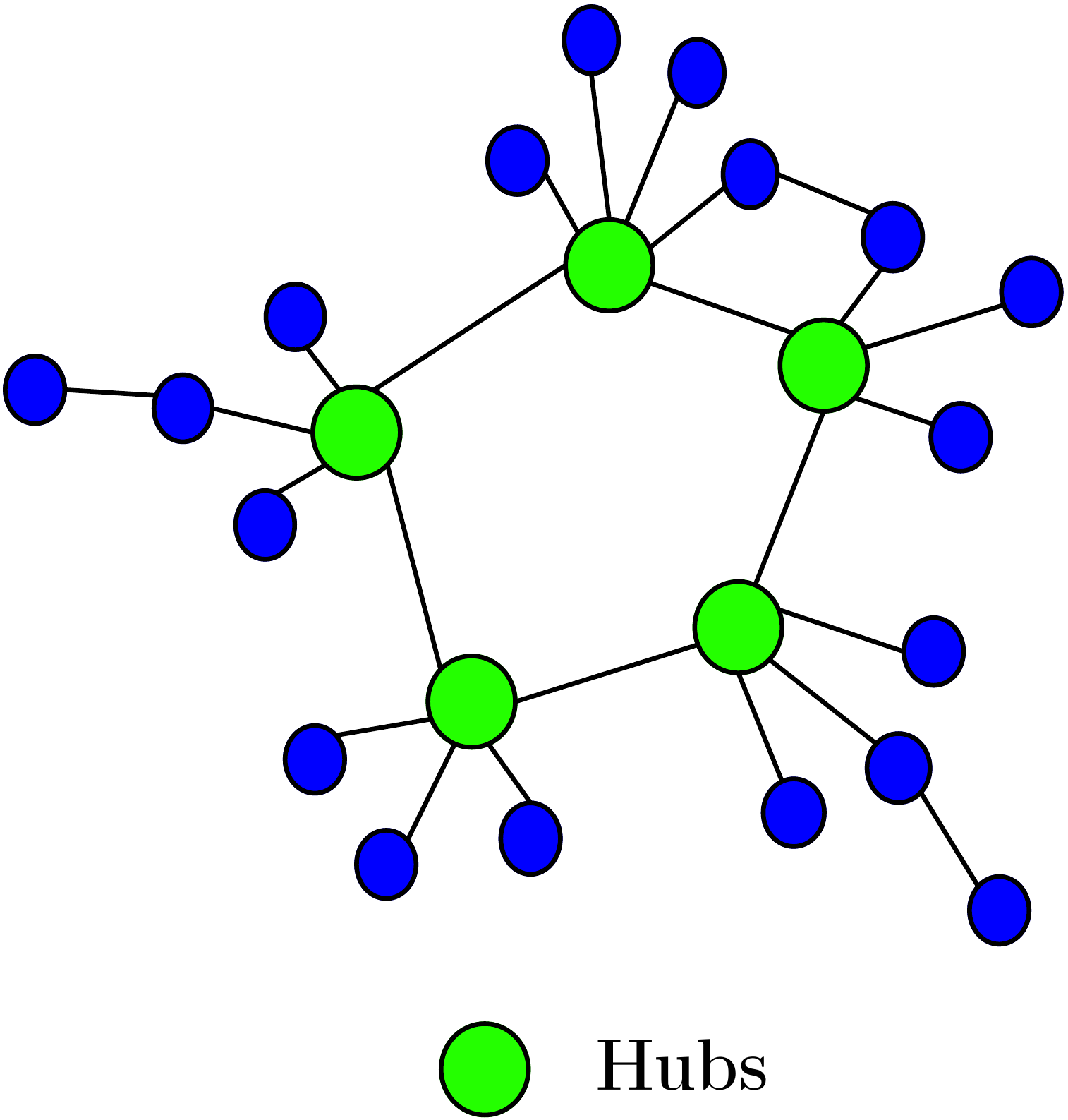}&
\includegraphics[width=0.45\linewidth, height=0.15\textheight]{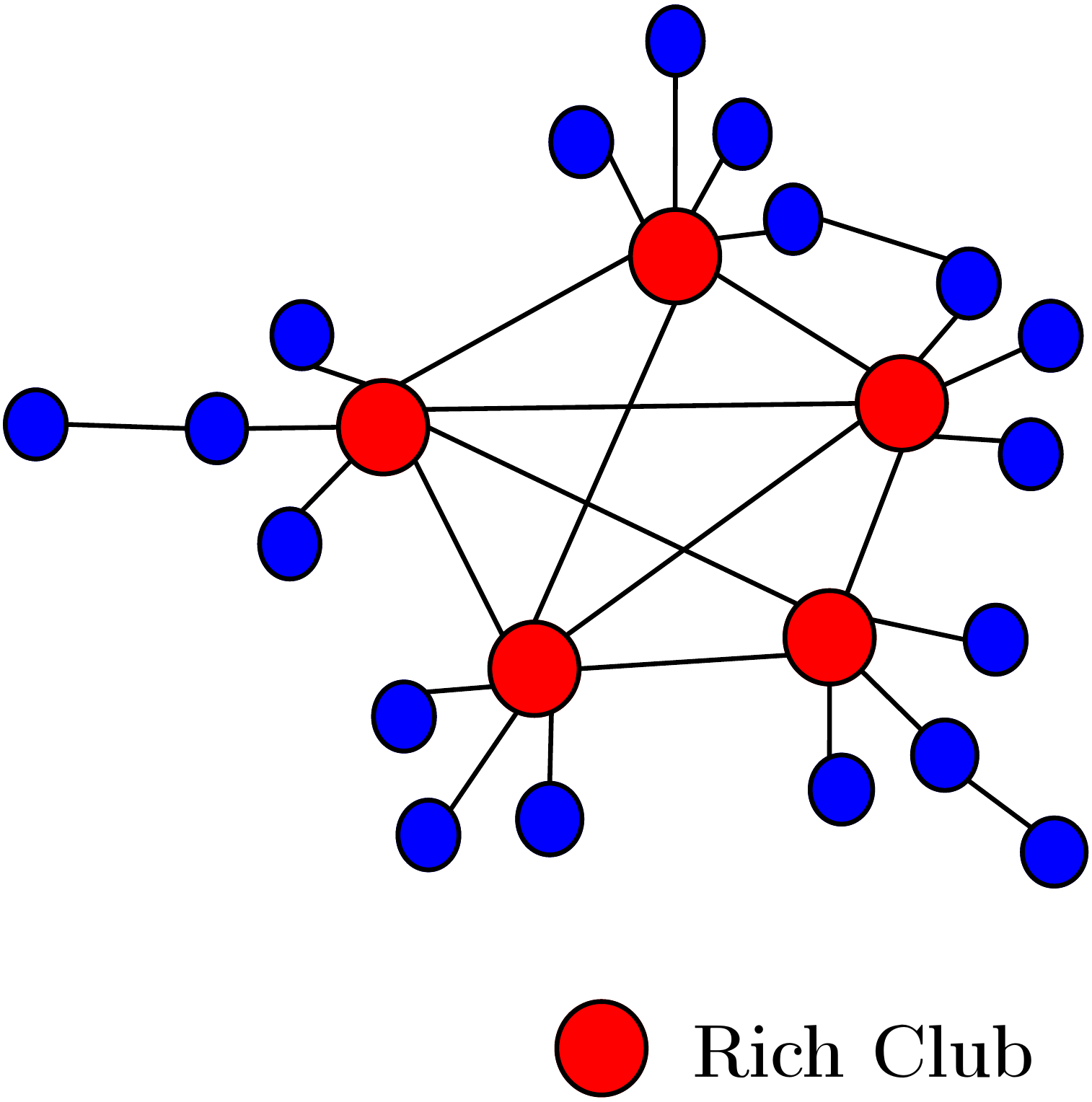}\\
\mbox{(a)} & \mbox{(b)}\\
\end{array}$
\caption{(a) A network composed of weakly interconnected hub nodes (green) and (b) Hub nodes now form a densely interconnected rich club (nodes in red), consisting of $ 5 $ nodes with a degree of $ 4 $ or higher.}
\label{f2}
\end{minipage}
\end{figure}

\smallskip
\hspace{-0.5 cm}\textbf{Core periphery architecture}\\
The core-periphery structure can be formed with $ k $-core decomposition, a method aimed to partitioned a network in layers from the external to the central ones \cite{rombach}. A $ k $-core of the graph $ G $ is a maximal connected subgraph in which all nodes have the degree at least $ k $. Let $ G = (V, E) $ is a network with $ V $ set of nodes and $ E $ set of links. A subset $ H = (W, L|W) $ is a maximum subgraph where, set $ W $ is a core of order $ k $ if and only if the nodes belong to set $ W $ have the degree $ k $ in the subgraph $ H $ i.e., $ \forall n \in W : k_n^H \geq k $. The core of maximum order is also called the innermost or main core. The core number assigned to a node $ n $ is the highest order of a core that contains this node. Cores are nested according to degree i.e., if the degree of a node $ i $ is less than the degree of a node $ j $ then the subgraph $ H $ containing $ j $ will be subset of the subgraph $ H $ containing $ i $ ($ i < j \Rightarrow H_j \subseteq H_i $) \cite{core}. The $ k $-core decomposition of a European airline is shown in Fig. \ref{f1}. The identification of the nodes in the $ k $-core is very helpful to find congestion in the network.

\smallskip
\hspace{-0.5 cm}\textbf{Rich club phenomenon}\\
Rich club phenomenon is characterized when the hubs are on average more intensely interconnected than the nodes with smaller degrees., rich club as a function of degree $ k $ is defined as $ \Phi(k) = \frac{2E_{>k}}{N_{>k}(N_{>k}-1)} $ where, $ N_{>k} $ is the number of nodes with degree greater than $ k $, and  $ E_{>k} $ is the number of links connecting the $N_{>k} $ nodes. A sample network with and without rich club is shown in Figure \ref{f2}.

\section{Proposed Rewiring Strategies} \label{Sec3}
There are various local, global and intermediate level network properties that affect the structure of the networks. In rewiring process, a link is removed from one end and simultaneously that end will be added to the other node in the network. The rewiring process may be done randomly or by using some heuristics. Here, in this paper, the removal of the link is random but, addition of the link after rewiring process is based on some strategies. The parameters for rewiring are considered as degree, disassortativity, centrality measures, and core periphery architecture. The degree may be used for preferential attachment (PA) in the network. The probability $ \Pi_i $ that a node $ i $ will be selected through PA is defined as $ \Pi_i  = \frac{k_i}{\sum_j k_j}$. In a disassortative network, connection among hubs (or small degree nodes) are avoided, but, prefer the connection between hubs and lower degree nodes. Centrality measures are very helpful to find central and congested nodes in the network especially betweenness centrality (BC), closeness centrality (CC) and Eigenvector centrality (EC). The core periphery (CP) architecture helps to find central node in terms of degree and distance. The CP architecture can also be formed using the nodes which are part of $ k $-core and having maximum CC. By considering the importance of the above parameters, various rewiring strategies can be defined as follows.
\subsection{Disassortativity and Preferential Attachment (DPA) based rewiring strategy}
Disassortativity and preferential attachment (PA) are local parameters. In PA, a node $ j $ connects with another node $ i $ with probability $ \Pi_i $ in the existing network. In the rewiring strategy, the link $ e_{ij} $ between nodes $ i $ and $ j $ is removed if the node $ i $ shows assortativeness $ 0 < r_{deg}(i,j) \leq 1 $ with node $ j $. The first node, $ i $  will establish a new connection with node $ v $ by preferring the higher degree of the node $ v $ and normalized disassortativeness, $ \zeta_v $ of node $ v $ with its neighbors, $ Ne(v) $ in the network. The value of $ \zeta_v $ is dependent on the correlation, $ corr(v) $ of node $ v $ with its neighbors. The $ corr(v) $ measures maximum disassortativeness of a node $ v $ with its neighbor. If a node $ v $ is more disassortative to its neighbor then the selection probability will be increased. The probability $ \Pi_v^r $ of node $ i $ to attach with the node $ v $ may be defined as,
\begin{equation}
\Pi_v^r =  \frac{k_v}{\sum_u k_u}  \zeta_v\label{e5}
\end{equation}
Where, $ \zeta_v = \frac{corr(v)}{\sum_{n: n \in Ne(v)} r_{deg}(v,n)} $, $ corr(v) = min \mbox{ } r_{deg}(v,n), \forall {n: n \in Ne(v)} $, and the value of $ r_{deg}(v,n) $ is scaled from the range of $ [-1,1] $ to $ [0,2] $. 
\subsection{Disassortativity and Eigenvector Centrality (DEC) based rewiring strategy}
Here, rewiring process is a combination of local and global parameters. In the proposed strategy, a link $ e_{ij} $ between node $ i $ and node $ j $ is chosen randomly. The node $ i $ will remove a connection with node $ j $ if both the nodes show assortativeness. Further, node $ i $ will prefer to connect with a new node $ v $ if overall congestion at the node is lesser than other nodes in the network. As EC finds the impact of a node on the other nodes in the network, therefore, it is considered for rewiring. Higher the value of EC, lower the probability to get selected for a new connection. That is why, the node $ i $ will prefer to connect with a node $ v $ having lower EC and higher disassortativeness with its neighbors. The probability $\Pi_i^{EC}  $ of node $ i $ to attach with node $ v $ may be defined as,
\begin{equation}
\Pi_i^{EC} =  (1-\hat{x}(v)) \zeta_v \label{e6}
\end{equation}
Where, $ \hat{x}(v) $ is Eigenvector centrality of node $ v $ and $ \zeta_v $ is same as explained in Eq. \eqref{e5}.
\subsection{$ K $-core (using Degree) and Betweenness Centrality (DKBC) based rewiring strategy}
This strategy is a combination of meso level and global parameter. The network is layered into $ k $-cores on the basis of the degree of nodes \cite{rombach}. In such structure, the nodes in the inner cores are more central than the nodes in the periphery. Hence, nodes with higher $k$-core value (i.e., inner core) are preferred for the connection in the rewiring process. After that, nodes in the inner cores are ranked according to BC of the nodes in the network. If a node lies in the shortest path with high BC then it may be congested early. Hence, in the proposed strategy, a node with low BC is chosen for rewiring. In the proposed rewiring strategy, first, a link, $ e_{ij} $ between the nodes, $ i $ and $ j $ is removed randomly. Then, the node $ i $ will be connected with a node $ v $ if it has higher $ k $-core value i.e., $ core(v) > core(i) $ and having less BC, i.e, $ g(v) < g(i) $. Where, $core(v) $, $ core(i) $, $g(v)$ and $ g(i) $ represent $ k$-core score of node $ v $, $ k$-core score of node $ i $, BC of node $ v $ and BC of node $ i $ respectively. The detailed description is presented in Algorithm \ref{al2}.
\begin{algorithm}[htb!]
\begin{algorithmic}[1]
\STATE \textbf{Input:}  Initial Network ($ G $), BC ($ g $).
\STATE \textbf{Output:} Rewired Network ($ G' $).
\STATE Store core number of each node $ n \in V $ in $ core[n]$ 
\STATE Remove a link $ e_{ij} $ randomly. 
\STATE Select a node $ v $ randomly.
\IF{$ core(v)  >  core(i) $   AND  $ g[i]  >  g[v] $ }
\STATE Add a link $ e_{iv} $ between the nodes $ i $ and $ v $.
\ENDIF
\\ return $ G' $
\end{algorithmic}
\caption{$ K $-core (using Degree) and Betweenness Centrality (DKBC) based rewiring strategy}
\label{al2}
\end{algorithm}    
\subsection{$ K $-core (using Closeness centrality), Disassortativity and Betweenness Centrality (CKDBC)  based rewiring strategy}
Nodes in the innermost core are more central (in terms of distance) from all other nodes in the network \cite{rombach}. Hence, the $ k $-core is computed on the basis of the closeness of a node from all other nodes in the network. If the nodes are closest to all other nodes then they will be a part of the innermost core. Based on the value of closeness centrality (CC) of nodes, the network is divided into multiples cores. The total number of core is kept same as the score of $ k $-core through the degree of nodes. The interval of each core depends on the difference of the maximum and minimum values of CC of nodes. Algorithm \ref{al3} provided detail description for the generation of $ k $-core structure. The nodes in the innermost core have higher CC hence, they can rewire their connections to reduce congestion in the network. A node $ i $ is chosen randomly from the innermost core and its connection with a node $ j $ is removed if the nodes $ i $ and $ j $ show assortativeness and the node $ j $ also holds higher BC values. A new node $ v $ is chosen if it shows disassortativeness with the node $ i $ and having less BC value. The detailed description is provided in Algorithm \ref{al4}.

\begin{algorithm}[htb!]
\begin{algorithmic}[1]
\STATE \textbf{Input:}  Network ($ G $).
\STATE \textbf{Output:} core[G]. 
\STATE Find CC of each node and store it into a set $ \mathcal{C} $
\STATE$ min  \leftarrow $  minimum($ \mathcal{C}) $ /* Assign minimum value of $ \mathcal{C} $ into $ min $ */
\STATE$ max  \leftarrow $  maximum($ \mathcal{C}) $ /* Assign maximum value of $ \mathcal{C} $ into $ max $ */
\STATE Find the main core of the network and save into $ numCore $.
\STATE Calculate interval of the cores, $ CI  \leftarrow \frac{(max-min)}{numCore} $.\\
/* Assign core index to each node $ i $ present in the set $ V $ and save into $ core $ */
\FOR{ $ i $ in $ V $}
\STATE index $ \leftarrow $ $ \frac{\mathcal{C}[i] -min}{CI} $.
\STATE Assign core index to node $ i $ and append into $ core $.
\ENDFOR
\\ return $ core $
\end{algorithmic}
\caption{Formation of core through closeness centrality, $ \mathcal{C} $ of the nodes}
\label{al3}
\end{algorithm}

\begin{algorithm}[htb!]
\begin{algorithmic}[1]
\STATE \textbf{Input:}  Initial Network ($ G $), BC ($ g $), assortativity coefficient ($ r_{deg} $), $ core $.
\STATE \textbf{Output:} Rewired Network ($ G' $). 
\STATE Pick a node $ i $ randomly from the innermost core.\\
/* Find the product of assortativity coefficient of each node $ n $ with node $ i $ and BC ($ g(n) $) and save into $ prod $ */  
\FOR{ $ n $ in $ V $}
\STATE $ prod(n) $ $ \leftarrow  (r_{deg}(i,n) + 1) * g(n) $
\ENDFOR
\STATE Select a node $ j $ randomly with probability proportional to $ prod $ and remove the link $ e_{ij} $.
\STATE Select a node $ v $ randomly form the set $ V $
\IF {$prod(v)   < random() $} 
\STATE Add a connection between the node $ i $ and the node $ v $
\ENDIF\\
return $ G'$
\end{algorithmic}
\caption{$K$-core (using Closeness centrality), Disassortativity and Betweenness Centrality (CKDBC)  based rewiring strategy}
\label{al4}
\end{algorithm}

\section{Traffic Model}\label{Sec4}
Network capacity and traffic load can be calculated by using a suitable traffic flow model. The capacity depends upon the influence of a node in the network and it can be found by using various centrality measures. The model should handle dynamic traffic as well as the load of the network by enhancing network capacity. The dynamics of the traffic flow model consists of the following steps
\begin{itemize}[\textendash]
\item At each time step, every node generates $ \lambda $ packets and the size of the network is $ N $. Hence, the total number of generated packets are $ \lambda  N $.
\item A node $ i $ may deliver at most $ C_i $ packets towards its destination node through the shortest path. The network capacity is the sum of the capacities of all the nodes which appear in the network and defined as $ \sum_{i=1}^N C_i $.  
\item The packets are stored in a queue at the nodes and processed on the basis of first-in-first-out (FIFO). 
\item At each time step, some new packets are generated and once they reach their destinations then they are removed from the system.
\item To maintain the free flow state, we need to maintain the balance between the generated and delivered packets.
\end{itemize}
\subsection{Theoretical Analysis}
Most of the real world networks are time-varying and evolving. The concept of evolving network was proposed by Barabasi and Albert. The Barabasi-Albert (BA) model helps to generate random scale-free networks via preferential attachment in terms of degrees of the nodes in the network. The network has variable data capacity for an individual node. Each node is endorsed by its neighboring nodes. If these neighbors are already congested then it will cause an increase in load at that neighbor node. For the network designed through BA model, it was shown that the BC is related to the degree via the relation $ g \sim k^{\frac{\alpha-1}{\eta-1}} $ \cite{goh2001} where, $ \alpha $ and $ \eta $ are power law exponents for degree and BC respectively. Therefore, larger degree nodes (hubs) tend to influence other nodes in communications by a greater extent, thus, making the hubs a part of the shortest paths. Congestion usually starts from the hub nodes. Hence, a rational capacity allocation scheme is required for the hub nodes in order to reduce data traffic in the network. Eigenvector centrality ($ EC $) is used to define the impact of the node on the other nodes in the network hence, it is considered in the computation for the capacity of a node. Betweenness centrality $ BC $ is used to find the involvement of the node in the shortest path of the respective user. Therefore, data forwarding capacity, $ C_i $ of a node $ i $ may be allocated as,
\begin{equation}
C_i^t = \beta (\hat{x}(i) + g(i)) N. \label{e1}
\end{equation}
Where, $ \hat{x}(i) $ is EC of a node $ i $ and $ g(i) $ is the $ BC $ of node $ i $. The term, $ \beta $ is a considerably modest fractional value and is a controlling parameter for capacity of the nodes. Packets are forwarded through the shortest paths from the source node to the destination node in the network. Therefore, the probability to pass through a node $ i $ is provided as $ \frac{g(i)}{\sum_{j} g(j) } $. At each time step, average number of incoming packets, $ Q_i^t $ at node $ i $ can be represented as,
\begin{equation}
Q_i^t =  \lambda \mathcal{D} N  \frac{g(i)}{\sum_{j} g(j)}. \label{e2}
\end{equation}
Where $ \mathcal{D} $ is the average shortest path of the network. The term, $ \lambda $ is a small fractional value and is a controlling parameter for $ Q_i^t $ for the node $ i $. With the increase of $ \lambda $ value, the system enters into a phase transition from a free flow state to a congested state. The point of the phase transition is known as critical point and the value of $ \lambda $ is considered as critical packet generation rate \cite{onset}, $ \lambda_c = \frac{C_{max} (N-1)}{g(max)}$. The capacity of a node increases with the increment of the value  $ \beta $. The sum of the capacities $ C_i $s of each node $ i $ is known as the capacity $ C $ of the network. Similarly, the sum of the incoming packets, $Q_i $s at each node $ i $  is termed as the traffic load, $ Q $, of the network. If the number of incoming packets exceeds the traffic capacity of the network then the system will be in the congested state otherwise, it will remain in the free flow state. In the free flow state, all the generated packets will be delivered and there will be no accumulation of packets in the network. However, in a congested state, the number of accumulated packets within the network increases with time. The capacity of the network depends on the capacity of the individual node. Hence, an optimal network structure can improve the performance (or throughput) of the network. To analyze throughput of the network, we need to find out an average of generated packets, $ \lambda N $ and delivery capacity $ C^t $ of the network at time $ t $. The total number of packets in the queue is $ L^t $ at time $ t $. From the theoretical perspective, $ L^{t+1} $ is equal to the sum of the total number of packets in the queue at time $ t $ and $ Q^{t+1} $ minus $ C^{t+1} $ and it can be expressed as,
\begin{equation}
L^{t+1} = L^{t} + Q^{t+1} - C^{t+1}
\end{equation}
The value of $ L^t $ will decide the total number of packets stored in the queue at time $ t $. For large $ t $, the value of $ L $ will be higher. A network structure will be considered as optimal if we get a lower value of $ L^t $ than the other network structures with the same number of nodes, $ N $ and average degree $ \langle k \rangle $. The proposed rewiring strategies consider congestion in the network. Hence, we will get a lower value of $ L^t $ than the original network at time $ t $. 

\section{Results and Analysis}
Since, most of the networks are time varying, we chose BA model to generate a scale-free (SF) network. In the SF network, the data forwarding capacity of the network is curtailed by various small degree nodes. Hence, we have proposed some rewiring strategies in Section \ref{Sec3} to minimize congestion and to enhance the performance (in terms of capacity) of the network. We have used a rational allocation of capacity distribution in Section \ref{Sec4} to avoid accumulation of the packets on the hub nodes in the network. The experimental setup consists of generated scale free networks and real data-sets as shown in Table \ref{tab3}. CSA dataset consists of social relationships (Facebook, Leisure, Work, Co-authorship, Lunch) between employees of computer science department at Aarhus. In EuAir,  nodes correspond to different airlines in Europe and the link shows the connection between them. Karate is a social network of friendships between members of a karate club at a US university. 

\begin{table}[!htb]
\caption{Experimental Setup}
 \center
 {
 \begin{tabular}[1]{|l| c|c|}
\hline
&Network & Parameters  \\ [0.5ex]
\hline\hline
Synthetic & Scale-free network & $ m_0 = 5 $\\
Network &(using BA model)  & $ \langle k \rangle = 4 $\\
& & $ N = 5 \times 10^2 $ to $ 2 \times 10^3 $\\[0.25ex]\hline
Real Network & CSA & $ N = 61 $, $ |E| = 620 $  \\
&EuAir & $ N = 450 $, $ |E| = 3588 $ \\
&Karate & $ N = 34 $, $ |E| = 64 $\\[0.25ex]\hline
\end{tabular}
\label{tab3}
}
\end{table}

The real world networks, like communication networks are large scale and heterogeneous in degrees, which then enable them to maintain the power law in degree distribution. Hence, the proposed rewiring strategies retain the degree distribution of the networks after each rewiring operation unlike some exisitng rewiring strategies \cite{jiang2014,gong,rateeqn}. Further, we analyzed the efficacy of the proposed rewiring strategies by studying various network properties on the synthetic network and on the empirical data-sets.

\subsection{Performance Evaluation}
The network performance is affected by various parameters, critical packet generation rate $ \lambda_c $, maximum betweenness centrality $ g(max) $, node utilization probability $ U_k $ and rich-club coefficient (RC). Hence, all these parameters are considered as an indicator of performance evaluation. 

\begin{figure}[htb]
\begin{center}
\includegraphics[width=\linewidth, height=4.5 in]{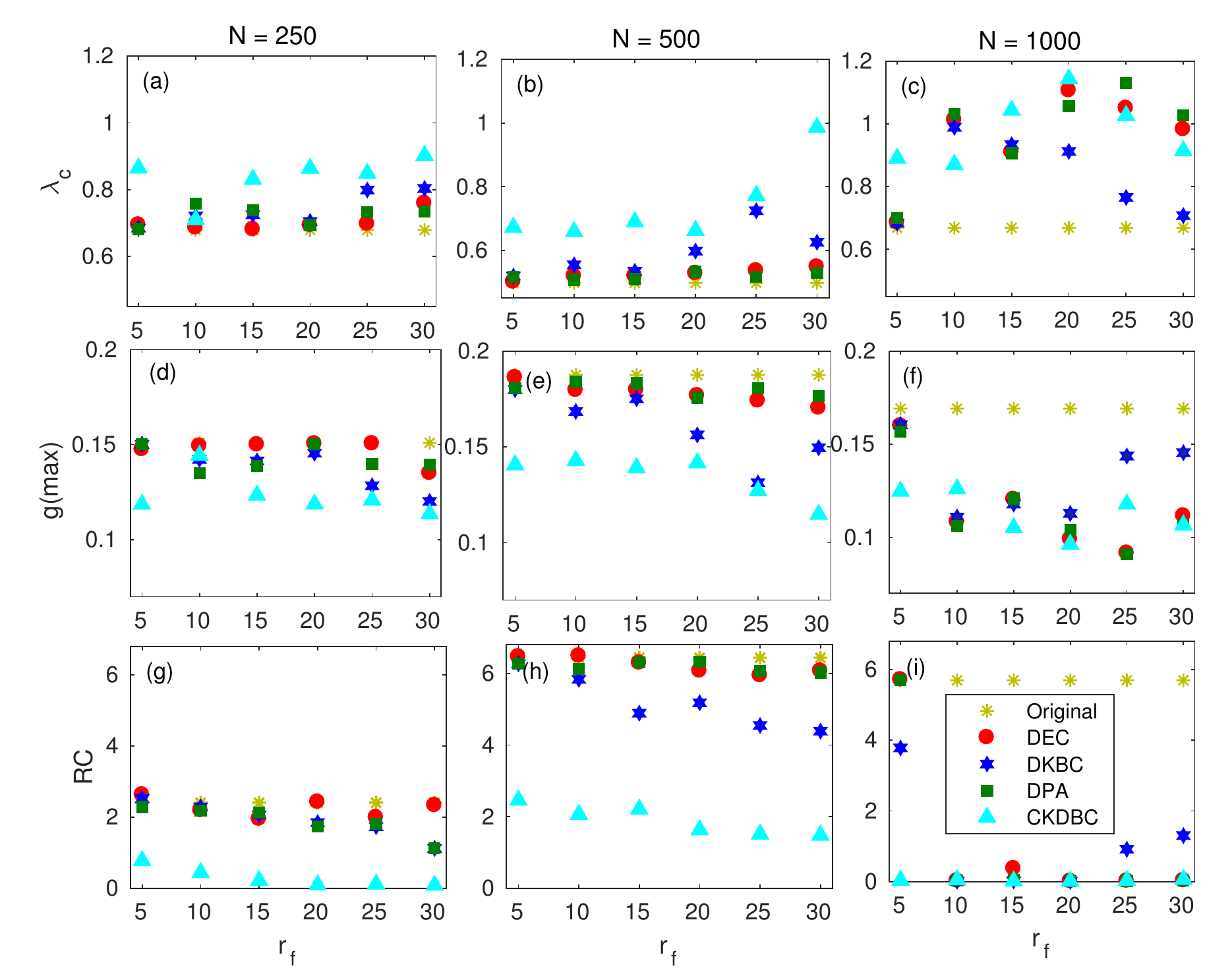} 
\caption{The evaluation of network parameters, $ \lambda_c $ (a-c), $ g(max) $ (d-f) and $ RC $ (g-i) for each rewiring strategy when the size of the network $ N = 250 $ (a,d,g), $ N = 500 $ (b,e,h) and $ N = 1000 $ (c,f,i). The X-axis represents the fraction of rewired links ($ r_f $). Each result value is the average of $ 10 $ independent realizations.}
\label{f3}
\end{center} 
\end{figure}

According to the traffic flow model, discussed in Section \ref{Sec4}, network congestion occurs beyond the critical packet generation rate $ \lambda_c $, but below it network traffic is free flow. The value of $\lambda_c$ is proportional to network capacity ($C$) inversely proportional to the BC of the node with maximum BC value, $g(max)$. Rich club phenomenon characterized when the hubs are on average more intensely interconnected than the nodes with smaller degrees. Presence of rich club increases load and congestion on the connecting link between two hubs. Hence, $ RC $ has a negative impact on capacity of the network. All these network parameters ($ \lambda_c, g(max) $ and $ RC $) are plotted for increasing value of $r_f$ (Fig. \ref{f3}) for varying size of the network ($ N = 250, N = 500 $ and $ N = 1000 $). We can infer that all the rewiring strategies give a higher value of $ \lambda_c $ and lower value of maximum BC, $ g(max) $ than the original network. The proposed rewiring strategies remove a link randomly from a randomly selected node. There exists a small fraction of links whose removal will affect the structure of the network drastically. The probability of selection of these nodes is very less. For that reason, rewiring of a large fraction of links is not much effective in the enhancement of the network performance. However, rewiring a link from a highly congested node will improve the network capacity and reduce congestion in a considerable manner. Rich club of the network is calculated by using rich club coefficient $ \phi(k) $ of the nodes with degree $ k $, and is written as $ \frac{k \phi(k)}{N} $. The proposed rewiring strategies take care of congestion and offer the lower value of RC than the initial network. The rewiring strategy based on the formation of $k$-core architecture through closeness centrality (CKDBC) provides lowest value of $ g(max) $ and $ RC $. Therefore, we get highest value of $ \lambda_c $ and is comparatively desirable for varying size of the scale free network (using BA model).
 
\begin{figure}[htb]
\begin{center}
$\begin{array}{ccc}
\includegraphics[width=0.33\linewidth, height=1.5 in]{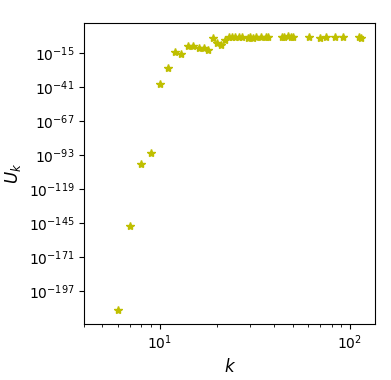} &
\includegraphics[width=0.33\linewidth, height=1.5 in]{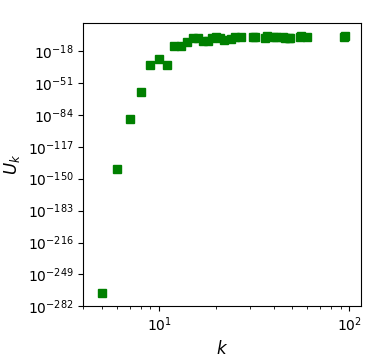} &
\includegraphics[width=0.33\linewidth, height=1.5 in]{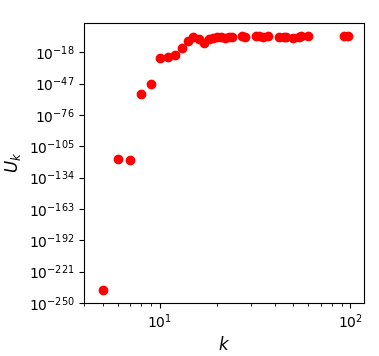}\\
\mbox{(a) Original Network} & \mbox{(b) DPA} & \mbox{(c) DEC} \\
\includegraphics[width=0.33\linewidth, height=1.5 in]{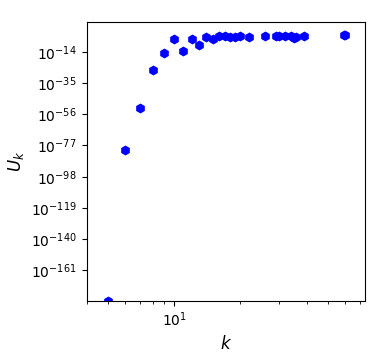} &
\includegraphics[width=0.33\linewidth, height=1.5 in]{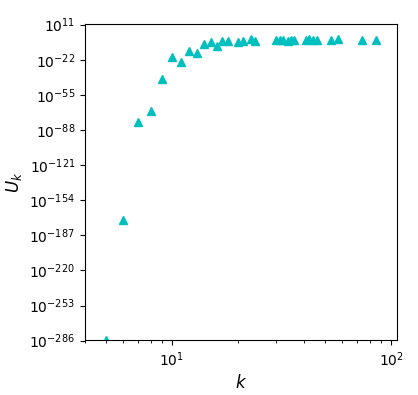}\\
\mbox{(d) DKBC} & \mbox{(e) CKDBC}
\end{array}$
\caption{Node utilization probability, $ U_k $ versus node degree, $ k $ for the original network as well as for the network designed through different rewiring strategies when $ r_f = 15\%$ and $ N = 1000 $.}
\label{f4}
\end{center} 
\end{figure}

In a communication network, few hub nodes are loaded with traffic and are more prone to the congestion. In the proposed traffic flow model, we have assigned more capacity to such nodes. However, the smaller value of maximum node utilization probability, $ U_{max} $ is desirable for improvement in network capacity. In Figure \ref{f4}, we used a log-log plot to show the node utilization probability $ U_k $ and the degree of the node $ k $. In the original network \ref{f4}(a), the traffic intensity on the hub nodes are much higher than the rest of the network. After rewiring of the network through various strategies, traffic of the hub nodes is redistributed among rest of the nodes in the network.  This results into the reduction in the value of $ U_{max} $ in such order: Original Network $ > $ DKBC $ > $ DEC $ > $ CKDBC $  > $ DPA. 

\begin{figure}[htb]
\begin{center}
\includegraphics[width=\linewidth, height=2 in]{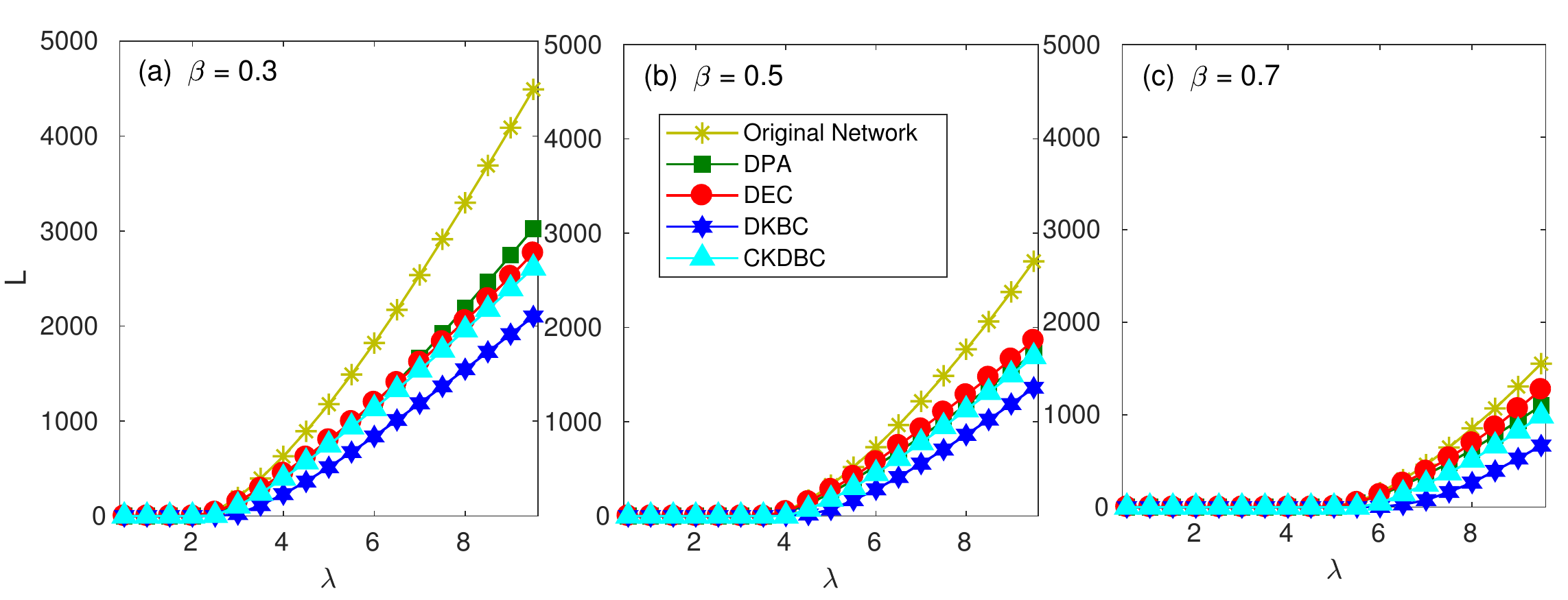} 
\caption{The X-axis represents the packet generation rate ($ \lambda $) and the Y-axis plots total number of accumulated packets $ L $ for different value of controlling parameters of capacity $ \beta $; (a) $ 0.3 $, (b) $ 0.5 $ and (c) $ 0.7 $ for each rewiring strategy. Each result value is the average of $ 10 $ independent realizations.}
\label{f5}
\end{center} 
\end{figure}

Figure \ref{f5} shows total number of undelivered packets $ L $ versus packet generation rate $ \lambda $ for different values of controlling parameters $ \beta $ for the network designed through different rewiring strategies. We analyze that, for
all the considered cases $ L $ is approximately zero for smaller value of $ \lambda $. There is a critical value of $ \lambda $ after which $ L $ increases. We also noticed that $ \lambda_c $ increases with $ \beta $, which means that the enhanced network capacity alleviate congestion from the large degree nodes. Accordingly, network becomes more tolerant to congestion with delayed phase transition. The rewiring strategy based on $ k $-core decomposition through degree (DKBC) is more tolerant to congestion. The order of $ L $ is: $ L_{Original\mbox{   } Network} > L_{DPA},L_{DEC} > L_{CKDBC} > L_{DKBC} $.

\subsection{Analysis of Synthetic Networks and Real World Networks}
\begin{figure}[htb]
\begin{center}
$\begin{array}{cc}
\includegraphics[width=0.45\linewidth, height=2.5 in]{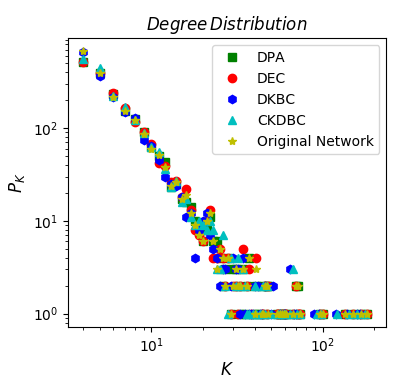} &
\includegraphics[width=0.45\linewidth, height=2.5 in]{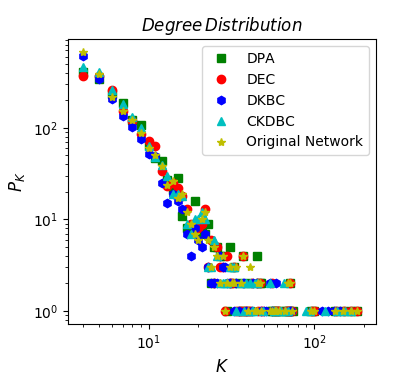} \\
\mbox{(a)} & \mbox{(b)} 
\end{array}$
\caption{Degree distribution of the network under different rewiring strategies when the size of the network, $ N = 2000$, average degree, $ \langle k \rangle = 4 $. The plot is log-log plot between degree $ k $ vs. the probability of the nodes having degree $ k, p(k) $ for (a) $ r_f = 5 \% $ and (b) $ r_f = 15\% $.}
\label{f6}
\end{center} 
\end{figure}

In most of the reported literature, the degree distribution has been considered as an important parameter for the study of the structure of the network. It has been proven that the degree distribution of the communication network is heterogeneous in nature and it affects the network performance in the case of dynamic traffic. The degree distribution after rewiring of $ 5\% $ of links in Figure \ref{f6}(a) and $ 15\% $ of the links (in Figure \ref{f6}(b)) is plotted through all the proposed rewiring strategies. As a result, we can infer that the rewired network is still following power law with little distortion in the degree of small degree node. 

Various network properties, e.g., $ g(max) $, $ \lambda_c $, assortativity coefficient ($ r_{deg} $), average clustering coefficient ($\langle C_lC \rangle$), average node coreness (ANC), average path length (APL), average node betweenness (ANB) are studied for scale-free networks (Table \ref{tab1}) and for real data-sets (Table \ref{tab2}). These are the properties of the network as a whole and not just of the individual node. All the rewiring strategies reduce $ g(max) $ and enhance $ \lambda_c $. The significance of the rewiring strategies is different. If someone is interested in the disassortative (or assortative) network then they will prefer CKDBC rewiring strategy as it provides a minimum value of $ g(max) $, the maximum value of $ \lambda_c $, most disassortative (or assortative) in nature and having least clustering coefficient. The CKDBC makes communication (scale-free) and transportation (EuAir) networks more disassortative while social network (CSA) is converted into a more assortative network. As the network through DPA strategy provides maximum value of $ \langle C_lC \rangle $ hence, more desirable for fast spreading process and robustness. ANC,  APL, and ANB of the network designed through all the rewiring strategies are almost equal (Table \ref{tab1}). But, the values of ANC, APL, and ANB are slightly changed for the empirical data-sets (Table \ref{tab2}).        
\begin{table}[!htb]
\caption{Topological measures for the network designed with various rewiring strategies  when $ N=500 $ and $ r_f= 10\% $.} 
 \center
 {
 \begin{tabular}[1]{|l| c| c| c| c| c|}
\hline
Measures &Original & DPA &  DEC & DKBC &  CKDBC  \\ [0.5ex]
& Network &&&&\\
\hline\hline
$ g(max) $ & $ 0.1876 $ & $ 0.1708 $ & $ 0.1703 $ & $ 0.1772 $ & $ 0.1551 $ \\[0.25ex]\hline
$ \lambda_c $&  $ 0.4977 $ &  $ 0.5466 $ & $ 0.5482 $ & $  0.5276 $ & $ 0.6053 $  \\[0.25ex]\hline
$ r_{deg} $ & $-0.0892 $ &$ -0.0787 $ & $ -0.0338 $ & $ -0.0581 $ & $ -0.2181 $ \\[0.25ex]\hline
$ \langle C_lC \rangle $ & $ 0.1091 $ &  $ 0.1227 $ & $ 0.0779 $ & $ 0.0897 $ & $ 0.0700 $ \\[0.25ex]\hline
ANC & $ 4.1640 $ &$ 4.2240 $ & $ 4.2440 $ & $ 4.2120 $ & $ 4.2700 $  \\[0.25ex]\hline
APL & $ 2.8895 $ & $ 2.9246 $ & $ 2.9595 $ & $  2.9311 $ & $ 2.9122 $ \\[0.25ex]\hline
ANB& $ 0.0038 $ & $ 0.0038 $ & $ 0.0039 $ & $ 0.0038 $ & $ 0.0038 $  \\[0.25ex]
\hline
\end{tabular}
\label{tab1}
}
\end{table}

\begin{table}[!htb]
\caption{Topological measures for the network designed with different data-sets; CSA, EuAir and Karate when $ r_f = 10\% $.} 
 \center
{
\begin{tabular}[1]{|l|c|c|c|c|c|c|}
\hline
Measures & Network &Original & DPA &  DEC & DKBC &  CKDBC  \\ [0.5ex]
& &Network &&&&\\
\hline\hline
& CSA & $ 0.3275 $ & $ 0.2069 $ & $ 0.2541 $ & $ 0.1781 $ & $ 0.1573 $ \\[0.25ex]
$ g(max) $ & EuAir & $  0.8474 $ & $ 0.7889 $ & $ 0.8257 $ & $ 0.7787 $ & $ 0.7698 $\\[0.25ex]
& Karate & $ 0.4376 $ & $ 0.3281 $ & $ 0.4317 $ & $ 0.4336 $ & $ 0.3833 $ \\[0.25ex] \hline

& CSA &  $ 0.0924 $ &  $ 0.0989 $ & $  0.2867 $ & $  0.2388 $ & $ 0.2904 $  \\[0.25ex]
$ \lambda_c $& EuAir & $ 0.2098 $ & $ 0.2254 $ & $ 0.2154 $ & $ 0.2284 $ & $ 0.2357 $\\[0.25ex]
& Karate & $ 0.2589 $ & $ 0.3454 $ & $ 0.2625 $ & $ 0.2613 $ & $ 0.2956 $ \\[0.25ex] \hline

 & CSA & $0.0046 $ &$0.0185 $ & $ -0.0026 $ & $  0.0169 $ & $ 0.0731 $ \\[0.25ex]
$ r_{deg} $ & EuAir & $ -0.5716 $ & $ -0.5419 $ & $ -0.5130 $ & $ -0.4934 $ & $ -0.6011 $ \\[0.25ex] 
& Karate & $  -0.4756 $ & $ -0.4559 $ & $ -0.4619 $ & $ -0.4084 $ & $  -0.5176 $ \\[0.25ex] \hline

$ \langle C_lC \rangle $ & CSA & $ 0.6733 $ &  $ 0.6440 $ & $ 0.6620 $ & $ 0.5124 $ & $ 0.5824$ \\[0.25ex]
& EuAir & $ 0.3994 $ & $ 0.3821 $ & $ 0.3585 $ & $ 0.3516 $ & $ 0.0875 $ \\[0.25ex] 
& Karate & $  0.5706 $ & $ 0.5298 $ & $ 0.5942 $ & $ 0.5444 $ & $ 0.3149 $ \\[0.25ex] \hline

ANC & CSA &  $ 4.5833 $ &$ 4.6166$ & $  4.7333 $ & $ 4.3833 $ & $ 4.4333 $  \\[0.25ex]
& EuAir & $ 2.0784 $ & $ 2.1764 $ & $ 2.1372 $ & $ 2.0784 $ & $ 1.8431 $\\[0.25ex] 
& Karate & $ 2.9117 $ & $ 2.8823 $ & $ 2.8529 $ & $ 2.9117 $ & $ 2.6470 $\\[0.25ex] \hline

APL &CSA & $ 3.1887 $ & $ 3.0056 $ & $ 3.1121 $ & $  2.7463 $ & $ 2.9751 $ \\[0.25ex]
& EuAir & $ 1.9725 $ & $ 2.0831 $ & $ 2.0698 $ & $ 2.3051 $ & $ 2.2102 $ \\[0.25ex] 
& Karate & $ 2.4082 $ & $ 2.2638 $ & $ 2.4171 $ & $ 2.4741 $ & $ 2.4884 $ \\[0.25ex] \hline

ANB& CSA & $ 0.0377 $ & $ 0.0346 $ & $ 0.0336 $ & $ 0.0301 $ & $ 0.0340 $  \\[0.25ex]
& EuAir & $ 0.0198 $ & $ 0.0221 $ & $ 0.0218 $ & $ 0.0266 $ & $ 0.0247 $ \\[0.25ex] 
& Karate & $ 0.0440 $ & $  0.0395 $ & $ 0.0443 $ & $ 0.0460 $ & $ 0.0465 $ \\[0.25ex]
\hline
\end{tabular}
\label{tab2}
}
\end{table}

\begin{figure}[htbp]
\begin{center}
\includegraphics[width=\linewidth, height=2 in]{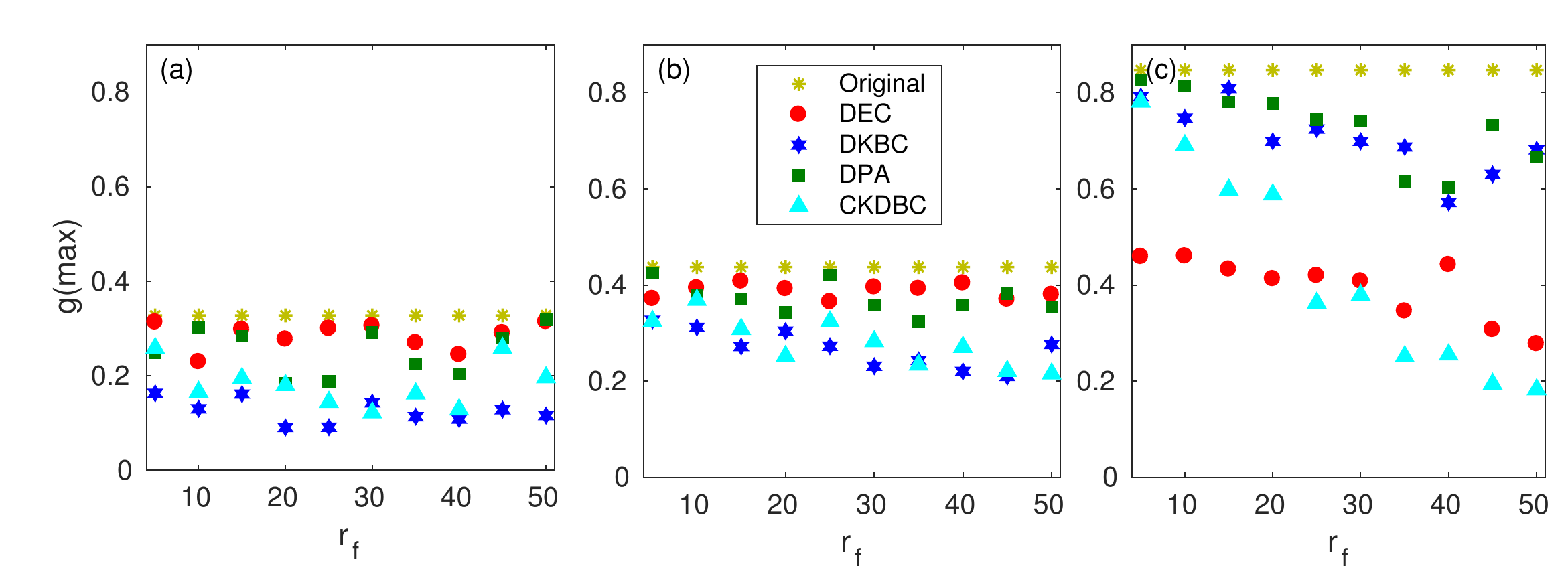} 
\caption{The X-axis represents the fraction of rewired links ($ r_f $) and the Y-axis plots the $ g(max) $ for each rewiring strategy for different datasets: (a) CSA \cite{CSA}, (b) Karate\cite{karate} (c) EuAir\cite{EuAir}. Each result value is the average of $ 10 $ independent realizations.}
\label{f7}
\end{center} 
\end{figure}

$ g(max) $ of all the data-sets are plotted against $ r_f $ in Fig. \ref{f7}. The value of $ g(max) $ obtained through all the rewiring strategies is lower than the original networks. CKDBC strategy outperforms than other strategies. DKBC performs better in all the data-sets except EuAir.
  
\begin{figure}[htb]
\begin{center}
\includegraphics[width=\linewidth, height=2 in]{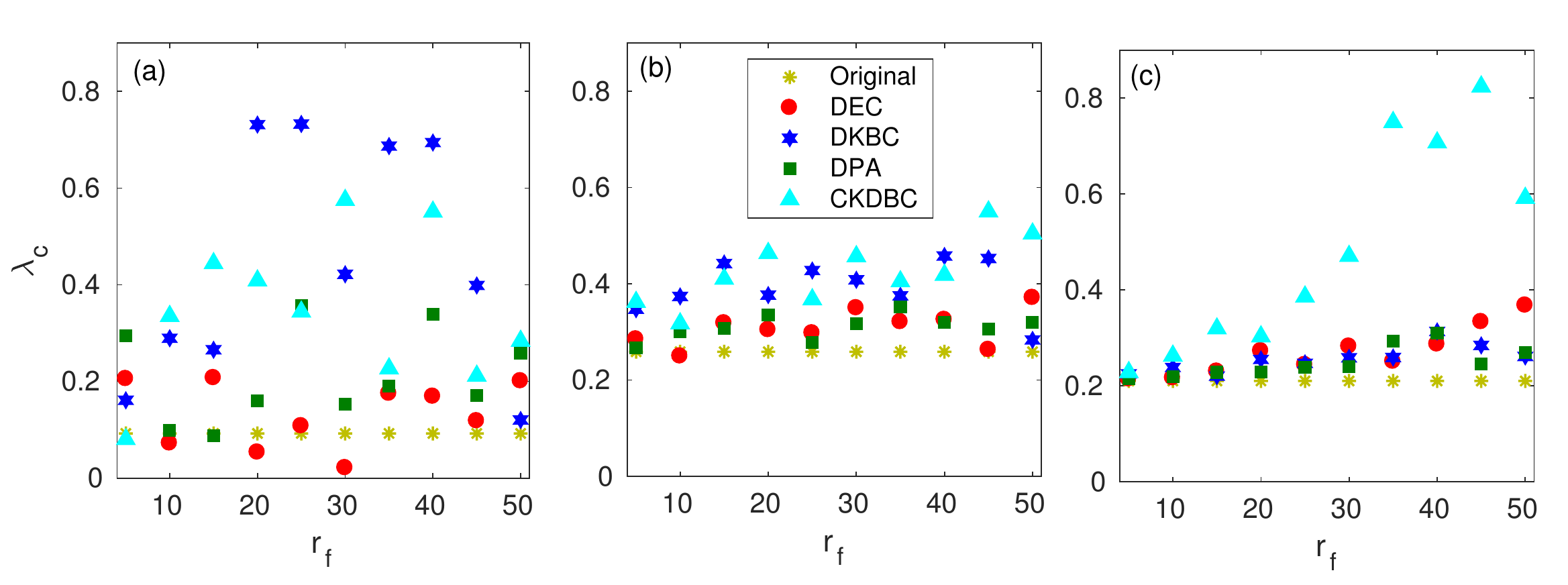} 
\caption{The X-axis represents the fraction of rewired links ($ r_f $) and the Y-axis plots critical packet generation rate, $ \lambda_c $ for each rewiring strategy for different data-sets. Each result value is the average of $ 10 $ independent realizations.}
\label{f8}
\end{center} 
\end{figure}

In Fig. \ref{f8}, among all the values of $ \lambda_c $, $ 95\% $ values of the proposed rewiring strategies for all the data-sets are higher than the $ \lambda_c  $ of the initial network. The DKBC routing strategy provides higher value of $ \lambda_c $ for friendship network (CSA and karate) while CKDBC outperforms for the transportation network (EuAir).

\begin{figure}[htb]
\begin{center}
\includegraphics[width=\linewidth, height=2 in]{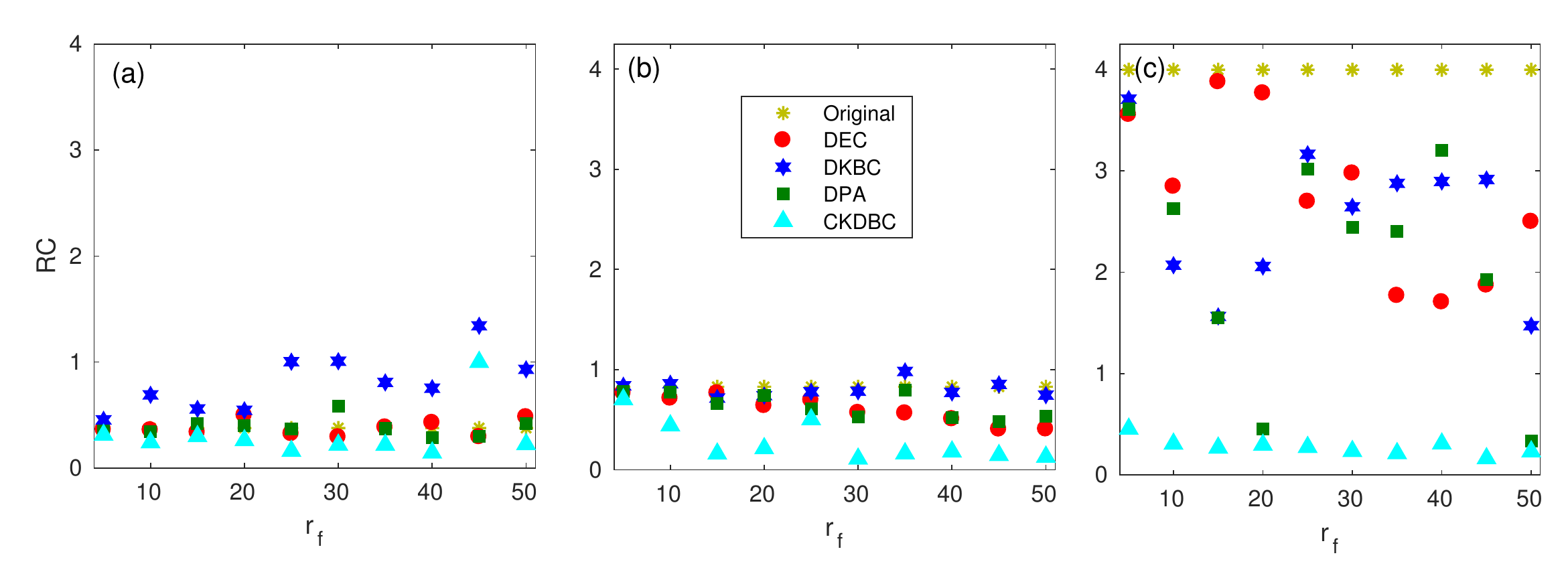} 
\caption{The X-axis represents the fraction of rewired links ($ r_f $) and the Y-axis plots rich club coefficient $ RC $ for each rewiring strategy for different data-sets. Each result value is the average of $ 10 $ independent realizations.}
\label{f9}
\end{center} 
\end{figure}

$ RC $ is plotted against $ r_f $(Fig. \ref{f9}). For all the data-sets, CKDBC strategy provides lower value of $ RC $ except a few $ r_f $s. For transportation (EuAir) network, $ RC $ for all the rewiring strategies are lower than initial datasets. 

\begin{figure}[htb]
\begin{center}
\includegraphics[width=\linewidth, height=2 in]{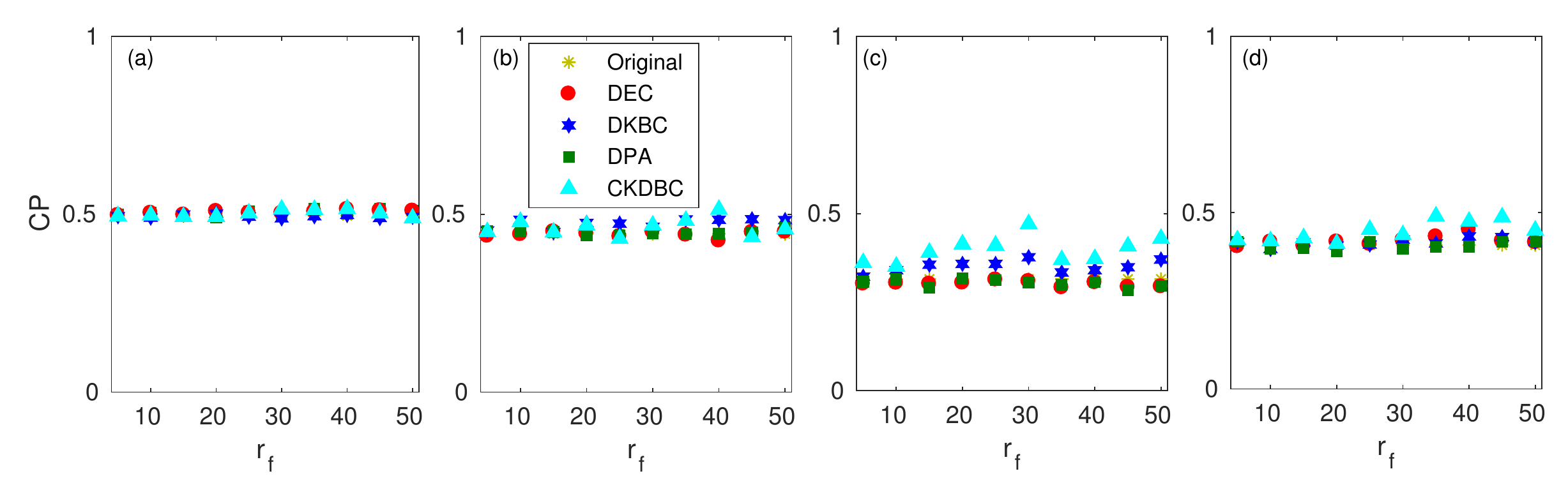} 
\caption{The X-axis represents the fraction of rewired links ($ r_f $) and the Y-axis plots core-periphery coefficient $ CP $ for each rewiring strategy for (a) the network designed through BA model when size $ N = 250 $ and (b-d) for different datasets. Each result value is the average of $ 10 $ independent realizations.}
\label{f10}
\end{center} 
\end{figure}

Core-periphery coefficient (CP) of the rewired (a) BA model and all the data-sets (b-d) are evaluated for different values of $ r_f $ (Fig. \ref{f10}). All the rewiring strategies maintain core-periphery architecture. Most of the $ CP $ values are higher than the value of initial data-sets. The performance of CKDBC rewiring strategy is better than others for all the data-sets. 

\section{Conclusions and Future Directions}
In this paper, efficient rewiring strategies are proposed to minimize congestion in the network by enhancing the critical packet generation rate, $ \lambda_c $ of the network. To mitigate traffic congestion, we have considered a combination of local, global and intermediate scale perspective for rewiring of the links in the network. As a result, it is found that the proposed rewiring strategies gives the higher value of $ \lambda_c $ and minimize betweenness centrality ($ g(max) $) along with rich club coefficient ($ RC $) while maintaining core-periphery architecture. Further, we analyzed various network properties on some empirical data-sets and find the same pattern as in scale free network (through BA model). Based on  the significance of the routing strategies, CKDBC is more desirable for data communication. While, DPA routing strategy can be preferred for fast spreading process and robustness.

In future work, we would like to expend our work to more realistic environment and those can be created using network simulators. Network congestion can also be studied for varying packet generation rate on different topologies of the networks.
 
 \section{References}
\bibliographystyle{splncs} 
\bibliography{NCC}
\end{document}